\documentclass[twocolumn,runningheads,natbib]{svjour2}
\smartqed  
\usepackage{graphicx}
\usepackage{aas_macros}
\usepackage{txfonts}
\journalname{Astrophysics and Space Science}
\begin{document}
\title{Long term spectral variability in the Soft Gamma-ray Repeater SGR~1900+14
}


\author{Andrea Tiengo        \and
        Paolo Esposito       \and
        Sandro Mereghetti    \and
        Lara Sidoli          \and
        Diego G\"{o}tz       \and
        Marco Feroci         \and
        Roberto Turolla      \and
        Silvia Zane          \and
        GianLuca Israel      \and
        Luigi Stella         \and
        Peter Woods
}


\institute{A. Tiengo \at
              INAF-IASF Milano, via Bassini 15, I-20133 Milan, Italy \\
              \email{tiengo@iasf-milano.inaf.it}           
           \and P. Esposito
       \at Universit\`a di Pavia, Dipartimento di Fisica Nucleare e Teorica and INFN-Pavia, via Bassi 6, I-27100 Pavia, Italy
           \and  P. Esposito \and S. Mereghetti \and L. Sidoli
       \at INAF-IASF Milano, via Bassini 15, I-20133 Milan, Italy
       \and D. G\"{o}tz
       \at CEA Saclay, DSM/DAPNIA/Service d'Astrophysique, \mbox{F-91191},
Gif-sur-Yvette, France
           \and M. Feroci
       \at INAF-IASF Roma, via Fosso del Cavaliere 100, I-00133 Roma,
Italy
           \and R. Turolla
       \at Universit\`a di Padova, Dipartimento di Fisica, via Marzolo
8, I-35131 Padova, Italy
           \and S. Zane
       \at Mullard Space Science Laboratory, University College
London, Holmbury St. Mary, Dorking Surrey, RH5 6NT, United Kingdom
           \and G.L. Israel \and L. Stella
       \at INAF - Osservatorio Astronomico di Roma,
 via Frascati 33, I-00040 Monteporzio Catone, Italy
           \and P. Woods
       \at Dynetics, Inc., 1000 Explorer Boulevard, Huntsville, AL 35806
}

\date{Received: date / Accepted: date}

\maketitle

\begin{abstract}
We present a systematic analysis of all the {\rm BeppoSAX} data of
SGR1900+14. The observations spanning five years show that the
source was brighter than usual on two occasions: $\sim$20 days after
the August 1998 giant flare and during the 10$^5$ s long X--ray
afterglow following the April 2001 intermediate flare. In the latter
case, we explore the possibility of describing the observed short
term spectral evolution only with a change of the temperature of the
blackbody component. In the only {\rm BeppoSAX} observation
performed before the giant flare, the spectrum of the SGR1900+14
persistent emission was significantly harder and detected also above
\mbox{10 keV} with the PDS instrument. In the last {\rm BeppoSAX}
observation (April 2002) the flux was at least a factor 1.2 below
the historical level, suggesting that the source was entering a
quiescent period.

\keywords{stars: individual (SGR~1900+14) \and stars: neutron \and
X--rays: stars} \PACS{97.60.Jd \and 98.70.Qy}
\end{abstract}

\section{Introduction}
\label{intro} The soft gamma-ray repeater (SGR) \mbox{SGR\,1900+14} was
discovered in 1979 through a series of short and soft gamma-ray
bursts  \citep{mazets79}. Many years later, its persistent pulsating
X--ray counterpart was discovered in the \mbox{2--10 keV} energy band
\citep{hur99}.
More recently, it was also detected in the hard X--ray range
(20--100 keV) with the INTEGRAL satellite, becoming the second SGR
with a measurable hard X--ray tail in the spectrum of its persistent
emission \citep{gotz06}.

The bursting activity of \mbox{SGR\,1900+14} is rather discontinuous
(see Figure \ref{fig:1}, bottom panel) and culminated on 1998 August
27 with the emission of a Giant Flare, when more than 10$^{44}$ ergs
of $\gamma$--rays were emitted in less than one second
\citep{hurley99gf,maz99,feroci01gf}. This was one of the three Giant
Flares detected up to now from three different SGRs and was
interpreted as a new evidence in favor of the {\it magnetar} model.
In this model \citep{thompson95,thompson96}, the SGRs and the
\linebreak Anomalous X--ray Pulsars (AXPs, another class of X--ray
sources with similar properties, \citealt{mereghetti02}) are
believed to be neutron stars powered by the decay of their extremely
intense magnetic field (\mbox{$B\sim10^{14}-10^{15}$ G}).

Here we present the analysis of the persistent emission of
\mbox{SGR\,1900+14} both in the soft and hard X--ray range and its evolution
across the Giant Flare and in relation to its bursting activity.

\section{Soft X--ray emission}
\label{sec:1}
\subsection{Observations and data analysis}
\label{sec1.1} We have analyzed the X--ray observations of
\mbox{SGR\,1900+14} performed with the {\rm BeppoSAX} satellite (see
Table \ref{tab:1}).
%
\begin{table}[t]
\caption{Log of the {\rm BeppoSAX} observations of SGR~1900+14}
\centering
\label{tab:1}       
\begin{tabular}{llll}
\hline\noalign{\smallskip}
Obs & Date & Instrument & Exp. time  \\[3pt]
\tableheadseprule\noalign{\smallskip}
A & 1997-05-12 & SAX/MECS & 46 ks\\
 &  & SAX/PDS & 20 ks\\
B & 1998-09-15 & SAX/MECS & 33 ks \\
 &  & SAX/PDS & 16 ks \\
C & 2000-03-30 & SAX/MECS & 40 ks \\
 & & SAX/PDS & 18 ks \\
D & 2000-04-25 & SAX/MECS & 40 ks \\
 & & SAX/PDS & 19 ks \\
E & 2001-04-18 & SAX/MECS & 46 ks \\
 & & SAX/PDS & 17 ks \\
F & 2001-04-29 & SAX/MECS & 58 ks \\
 & & SAX/PDS & 26 ks \\
G & 2002-03-09 & SAX/PDS & 48 ks \\
H & 2002-04-27 & SAX/MECS & 83 ks \\
\noalign{\smallskip}\hline
\end{tabular}
\end{table}

The spectra were extracted from
the MECS \citep{boella97mecs} and LECS \citep{parmar97} instruments
using circles with radii 4$'$ and 8$'$, respectively.
The background spectra were
extracted in all cases from nearby regions. Time filters were applied
to both the
source and background spectra to exclude the SGR bursts
detected during observation A, B and F.
The standard response matrices were used for
the MECS and LECS spectra.

\subsection{Spectral results}
\begin{figure}
\centering
  \includegraphics[width=0.9\textwidth,angle=90]{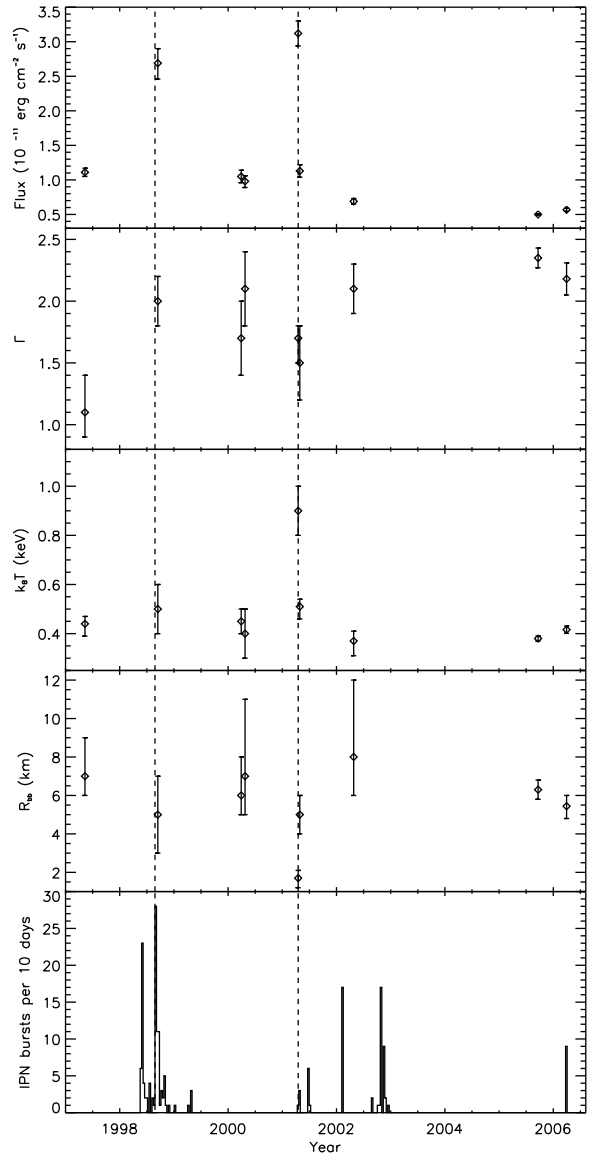}
\caption{Long term evolution of the 2--10 keV unabsorbed flux, the
spectral parameters (for an absorbed power-law plus blackbody model,
assuming N$_{\rm H}$=2.55$\times$10$^{22}$ cm$^{-2}$) and the burst
activity (as observed by interplanetary network) of
\mbox{SGR\,1900+14}. The vertical dashed lines indicate the times of
the Giant and Intermediate Flare (1998 August 27 and 2001 April 18,
respectively). We also plotted the 2005--2006 points obtained with
{\rm XMM-Newton} observations (Mereghetti et al. 2006, in
preparation).}
\label{fig:1}       
\end{figure}

We have first tried to fit the spectra with an absorbed power-law
model, but three observations  give unacceptable values of the
$\chi^2$ and structured residuals. For these observations, a good
fit is obtained with the addition of a blackbody component. Since
such a two-components model is typical of the magnetar candidates
\citep{woods04}, we have used this model to fit all the available
spectra, obtaining the results reported in Figure \ref{fig:1}. As
can be seen in  the upper panel, the flux varies by a factor $>$5,
with the highest values observed during observations B and E. These
two observations were taken shortly after extreme bursting events.
The former was performed 20 days after the Giant Flare, that was
followed by a $\sim$2 months period of enhanced X--ray flux
\citep{woods99b}. The latter started only 7.5 hours after the 2001
April 18 Intermediate Flare which had a fluence $\sim$20 times lower
than that of the Giant Flare \citep{feroci03}. The afterglow
following this bright burst is clearly visible during the {\rm
BeppoSAX} observation as a decrease in the X--ray flux, accompanied
by a significant softening of the spectrum \citep{feroci03}. In
addition to the afterglow analysis already reported by Feroci et al.
(2003), we have performed a time resolved spectroscopy of the
afterglow by dividing observation E into 5 time intervals. Although
the 5 spectra can be fitted by a variety of models, the spectral
evolution of the afterglow is well represented by an additional
blackbody component with fixed emitting area ($\sim$1.5 km, for a
source distance of 15 kpc) and progressively decreasing temperature
(\mbox{$k_BT$$\sim$1.3--0.9 keV}), that can be interpreted as due to
a portion of the neutron star surface heated during the flare.

Excluding the two observations taken after the exceptional explosive
events (B and E), the flux of \mbox{SGR\,1900+14} had a rather
constant value of $\sim$$10^{-11}$ erg cm$^{-2}$ s$^{-1}$ from 1997
to 2001. On the other hand a significantly lower flux level was seen
in the following observations. The flux decrease actually started
when the source was still moderately active (the flux in observation
H is at least 1.2 times lower than in all the previous quiescent
observations) and has been interrupted by a slight rise in
coincidence with the March 2006 burst reactivation, as shown by
recent {\rm XMM-Newton} observations (Mereghetti et al. 2006, in
preparation).

\begin{figure}
\centering
  \includegraphics[width=0.3\textwidth,angle=-90]{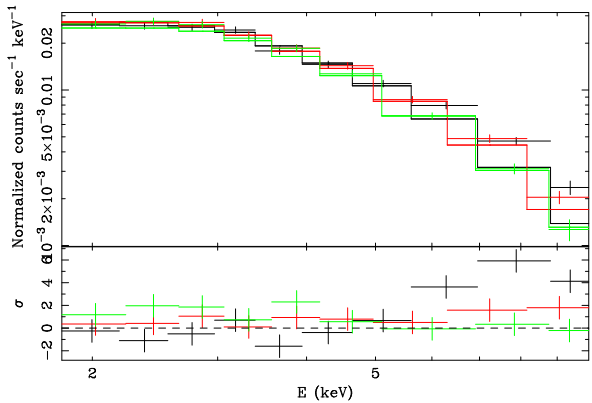}
\caption{{\rm BeppoSAX}/MECS spectra of observations A (black), C
(red) and D (green) simultaneously fit, with the same parameters, to
an absorbed power-law plus blackbody model. The data have been
rebinned graphically to emphasize the trend in the spectral
residuals of observation A.} \label{fig:res}
\end{figure}

Although the flux of the only pre-Giant Flare observation is
compatible with that of the quiescent post-flare observations taken
before 2002, its spectrum is significantly harder, as indicated by
the comparison of the photon indexes plotted in the second panel of
Figure \ref{fig:1}. The spectral change between observation A and
the following quiescent observations is well illustrated by the
clear trend of the residuals from the simultaneous fit of the
spectra of observations A, C and D (Figure \ref{fig:res}).

The blackbody parameters are compatible in all the available
observations, except for that taken during the afterglow of the
Intermediate Flare (observation E). This indicates that a constant
blackbody component with $k_BT$$\sim$0.4 keV and emitting area with \mbox{R$\sim$6-7 km} might be a permanent feature of the X-ray
spectrum of SGR\,1900+14.

\section{Hard X--ray emission} \label{sec:2}

\subsection{Detection with the PDS instrument}
To study the high energy emission from \mbox{SGR\,1900+14} we used
the {\rm BeppoSAX} PDS instrument, which operated in the 15-300 keV
range. The PDS instrument was more sensitive than INTEGRAL in this
energy band, but it had no imaging capabilities and therefore the
possible contamination from nearby sources must be taken into
account. The field of view (FoV) of the PDS instrument was
1.3$^{\circ}$ (FWHM) and the background subtraction was performed
through a rocking system that pointed to two 3.5$^{\circ}$ offset
positions every 96 s. In the case of \mbox{SGR\,1900+14}, the
background pointings were free of contaminating sources, as
confirmed by the identical count rates observed in the two offset
positions during each observation. The field of \mbox{SGR\,1900+14}
is instead rather crowded, with three transient sources, the X--ray
pulsars 4U~1907+97 \citep{giacconi71,liu00} and XTE~J1906+09
\citep{marsden981906}, and the black hole candidate XTE~J1908+94
\citep{intzand02}, located at angular distances of 47$'$, 33$'$ and
24$'$ from the SGR, respectively. The pulsations of the two pulsars
are clearly visible in the PDS data below 50 keV when they are
active, while XTE~J1908+94, if in outburst, is clearly visible in
the simultaneous MECS and LECS images and, being very bright, also
in the lightcurve collected by the All Sky Monitor (ASM) on board
the {\rm RossiXTE} satellite. We have found that at least one of
these contaminating sources was on in all the {\rm BeppoSAX}
observations except for the first one (see Table \ref{tab:2}). Thus,
only the 1997 observation (obs. A), during which a significant
signal was detected in the background subtracted PDS data, can be
used to study SGR\,1900+14 without the problem of known
contaminating sources.

\begin{table}[t]
\caption{Status of the three transient sources within the PDS field
of view during the {\rm BeppoSAX} observations. The presence of the
two X--ray pulsars (4U~1907+97 and XTE~J1906+09) is confirmed by the
presence of their pulsations in the PDS data, while that of the
black hole candidate (XTE~J1908+94) by its detection in the MECS and
LECS images and in the {\rm RossiXTE} ASM lightcurve.} \centering
\label{tab:2}       
\begin{tabular}{llll}
\hline\noalign{\smallskip}
Obs & 4U~1907+97 & XTE~J1906+09 &  XTE~J1908+94\\[3pt]
\tableheadseprule\noalign{\smallskip}
A & OFF & OFF & OFF\\
B & ON & ON & OFF \\
C & OFF & ON & OFF \\
D & ON & ON & OFF \\
E & OFF & ON & OFF \\
F & ON & ON & OFF \\
G & OFF & ON & ON \\
\noalign{\smallskip}\hline
\end{tabular}
\end{table}


We searched for the
SGR pulsation period (5.15719 s, as measured in the simultaneous
MECS data) in the PDS data, but the result was not conclusive,
giving only a 3$\sigma$ upper limit of 50\% to the pulsed fraction
of a sinusoidal periodicity, to be compared to the $\sim$20\% pulsed
fraction observed below \mbox{10 keV}.

\subsection{Spectral analysis}
\begin{figure}
\centering
  \includegraphics[width=0.35\textwidth,angle=-90]{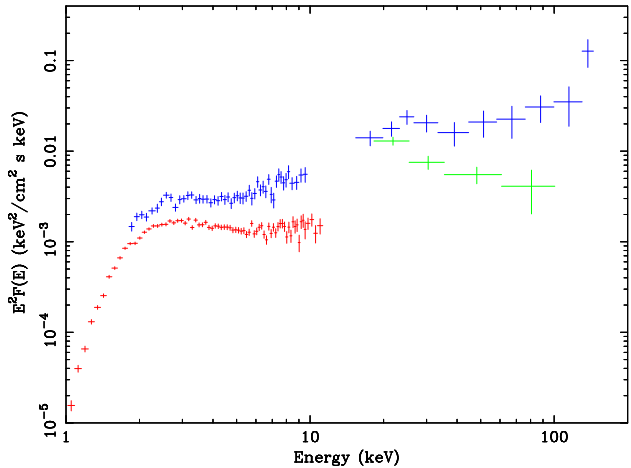}
\caption{Broad band spectra of \mbox{SGR\,1900+14} taken on 1997 May
12 (observation A) with {\rm BeppoSAX} (blue, both MECS and PDS
data), on 2005 September 20-22 by {\rm XMM-Newton} (red) and from
March 2003 to June 2004 with INTEGRAL.} \label{fig:2}
\end{figure}

The background subtracted PDS spectrum of observation A can be well
fit by a power-law with photon index \linebreak \mbox{$\Gamma=1.6\pm0.3$},
significantly flatter than that measured by INTEGRAL
(\mbox{$\Gamma=3.1\pm0.5$}, see Figure \ref{fig:2}) in 2003\,/\,2004. The
corresponding 20--100 keV flux is 6$\times$10$^{-11}$ erg cm$^{-2}$
s$^{-1}$, a factor $\sim$4 higher than during the INTEGRAL
observations, which confirms that before the Giant Flare
 the hard X--ray tail of \mbox{SGR\,1900+14} was
brighter.

The INTEGRAL spectrum was collected during $\sim$2.5 Ms of different
observations performed between March 2003 and June 2004, and thus it
represents the hard X--ray emission of \mbox{SGR\,1900+14} averaged
over that long time period. Therefore, its relation to the soft
X--ray spectrum can be studied only comparing the spectra taken by
other instruments in a similar time period, as shown for example in
Figure \ref{fig:2}. The PDS instrument, instead, being a high
sensitivity hard X--ray detector coupled to the MECS and LECS soft
X--ray cameras, gives us the chance to study the broad band spectrum
of \mbox{SGR\,1900+14} during a single observation. Fitting the
\mbox{1--150 keV} {\rm BeppoSAX} spectrum of observation A, we
obtain a good result ($\chi^2$=1.17 for 136 degrees of freedom)
simply extrapolating to higher energies the best-fit model found in
the soft X--ray range. In fact, a fit with an absorbed power-law
plus blackbody model gives the following parameters: photon index
$\Gamma=1.04\pm0.08$, blackbody temperature $k_BT=0.50\pm0.06$,
radius $R_{bb}=5\pm2$ km, and absorption $N_{\rm
H}=(1.8\pm0.5)\times10^{22}$ cm$^{-2}$.


%
%

\section{Conclusions} \label{sec:3}
We have studied the variability of \mbox{SGR\,1900+14}, both in the hard and
in the soft X--ray range, finding the following results:

\begin{itemize}

\item Except for the observations immediately following exceptional
flares, the flux level in soft X--rays was stable while the
source was moderately active and progressively decreased when
it entered a 3 years long quiescent period.

\item The Intermediate Flare of 2001 April 18 was followed by an
X--ray afterglow that can be successfully interpreted as due to the
heating of a significant fraction of the neutron star surface, that
then cools down in $\sim$1 day. This is consistent with the
interpretation of similar events in other magnetar candidates
\citep{woods2004}.

\item The soft X--ray spectrum during the only available pre-flare
observation was harder than in the following quiescent observations.
This is similar to what observed in SGR~1806--20, the only other SGR
that could be monitored before and after a Giant Flare \citep{mte05,rea05,tiengo05}.

\item Comparing the hard X-ray spectrum of \mbox{SGR\,1900+14} recently observed with
INTEGRAL to that observed with the PDS instrument in 1997,
 we find evidence for variations in flux and spectral slope.

\item The reduction of the X--ray tail in coincidence with the Giant Flare is supported
by the count rates detected in the PDS instrument above 50 keV
during the different {\rm BeppoSAX} observations, that indicate how
they significantly decreased already in the first post-flare
observation. Since the hard X--ray tail in the spectrum of
\linebreak \mbox{SGR\,1900+14} might contain most of its total
emitted energy, its variability in relation to the bursting activity
is a key point to try to understand the SGR emission processes.

\end{itemize}


\bibliographystyle{spmpsci}

\begin{thebibliography}{22}

\bibitem[\protect\citeauthoryear{Boella et al.}{1997}]{boella97mecs} Boella G., Chiappetti L., Conti G.
et al. \aaps, {\bf 122}, 327 (1997)

\bibitem[\protect\citeauthoryear{Feroci et al.}{2001}]{feroci01gf} Feroci M., Hurley K., Duncan R.C. et al. \apj, {\bf 549}, 1021 (2001)

\bibitem[\protect\citeauthoryear{Feroci et al.}{2003}]{feroci03} Feroci M., Mereghetti S., Woods P. et al. \apj, {\bf 596}, 470 (2003)

\bibitem[\protect\citeauthoryear{Giacconi et al.}{1971}]{giacconi71} Giacconi R., Kellogg E., Gorenstein P. et al. \apjl, {\bf 165}, L27 (1971)

\bibitem[\protect\citeauthoryear{G{\"o}tz et al.}{2006}]{gotz06} G{\"o}tz D., Mereghetti S., Tiengo A. et al. \aap,
  {\bf 449}, L31 (2006)

\bibitem[\protect\citeauthoryear{Hurley et al.}{1999a}]{hurley99gf} Hurley K., Cline T., Mazets E. et al. \nat, {\bf 397}, 41 (1999a)

\bibitem[\protect\citeauthoryear{Hurley et al.}{1999b}]{hur99} Hurley K., Kouveliotou C., Woods P. et al. \apjl, {\bf 510}, L107 (1999b)

\bibitem[\protect\citeauthoryear{in't Zand et al.}{2002}]{intzand02} in't Zand J.J.M., Miller J.M., Oosterbroek
T. et al. \aap, {\bf 394}, 553 (2002)

\bibitem[\protect\citeauthoryear{Liu et al.}{2000}]{liu00} Liu Q.Z., van Paradijs J. \& van den Heuvel
E.P.J. \aaps, {\bf 147}, 25 (2000)

\bibitem[\protect\citeauthoryear{Marsden et al.}{1998}]{marsden981906} Marsden D., Gruber D.E., Heindl W.A. et al. \apjl, {\bf 502}, L129 (1998)

\bibitem[\protect\citeauthoryear{Mazets et al.}{1999}]{maz99} Mazets E.P., Cline T.L., Aptekar R.L.
et al. Astronomy Letters, {\bf 25}, 628 (1999)

\bibitem[\protect\citeauthoryear{Mazets et al.}{1979}]{mazets79} Mazets E.P., Golenetskii S.V. \& Guryan Y.A. Soviet Astronomy Letters, {\bf 5}, 343 (1979)

\bibitem[\protect\citeauthoryear{Mereghetti et al.}{2002}]{mereghetti02} Mereghetti S., Chiarlone L., Israel G.L.
et al. in ``Neutron Stars, Pulsars, and Supernova Remnants", ed. W.
Becker, H. Lesch \& J. Tr{\"u}mper, [ArXiv: astro-ph/0205122] (2002)

\bibitem[\protect\citeauthoryear{Mereghetti et al.}{2005}]{mte05} Mereghetti S., Tiengo A., Esposito P. et al. \apj, {\bf 628}, 938 (2005)

\bibitem[\protect\citeauthoryear{Parmar et al.}{1997}]{parmar97} Parmar A.N., Martin D.D.E., Bavdaz M. et al. \aaps, {\bf 122},
309 (1997)

\bibitem[\protect\citeauthoryear{Rea et al.}{2005}]{rea05} Rea N., Israel G., Covino S. et al. The Astronomer's Telegram, {\bf 645} (2005)

\bibitem[\protect\citeauthoryear{Thompson \& Duncan}{1995}]{thompson95}
Thompson C. \& Duncan R.C. \mnras, {\bf 275}, 255 (1995)

\bibitem[\protect\citeauthoryear{Thompson \& Duncan}{1996}]{thompson96}
Thompson C. \& Duncan R.C. \apj, {\bf 473}, 322 (1996)

\bibitem[\protect\citeauthoryear{Tiengo et al.}{2005}]{tiengo05} Tiengo A., Esposito P., Mereghetti S. et al. \aap, {\bf 440}, L63 (2005)

\bibitem[\protect\citeauthoryear{Woods et al.}{1999}]{woods99b} Woods P.M., Kouveliotou C., van Paradijs J. et al. \apjl, {\bf 518}, L103 (1999)

\bibitem[\protect\citeauthoryear{Woods et al.}{2004}]{woods2004} Woods P.M., Kaspi V.M., Thompson C. et al. \apj, {\bf 605}, 378 (2004)

\bibitem[\protect\citeauthoryear{Woods \& Thompson}{2004}]{woods04}
Woods P.M. \& Thompson C. in ``Compact Stellar X-ray Sources", ed.
W.H.G. Lewin and M. van der Klis, [ArXiv: astro-ph/0406133] (2004)

\end{thebibliography}

\end{document}